\documentclass[12pt,english,number]{elsarticle}
\usepackage[T1]{fontenc}
\usepackage[latin9]{inputenc}
\usepackage{amsmath}
\usepackage{graphicx}
\usepackage{amssymb}
\usepackage{babel}
\usepackage{epstopdf}

\begin{document}

\title{Rich-club and page-club coefficients for directed graphs}

\author[label1]{Daniel Smilkov}

\author[label1,label2]{Ljupco Kocarev}

\address[label1]{Macedonian Academy of Sciences and Arts, Skopje, Macedonia \\
 Email: dsmilkov@cs.manu.edu.mk }

\address[label2]{ BioCircuits Institute, \\
 University of California San Diego, La Jolla, CA, USA \\
 Email: lkocarev@ucsd.edu }
\begin{abstract}
Rich-club and page-club coefficients and their null models are introduced
for directed graphs. Null models allow for a quantitative discussion
of the rich-club and page-club phenomena. These coefficients are computed
for four directed real-world networks: Arxiv High Energy Physics paper
citation network, Web network (released from Google), Citation network
among US Patents, and Email network from a EU research institution.
The results show a high correlation between rich-club and page-club
ordering. For journal paper citation network, we identify both rich-club
and page-club ordering, showing that {}``elite'' papers are cited
by other {}``elite'' papers. Google web network shows partial rich-club
and page-club ordering up to some point and then a narrow declining
of the corresponding normalized coefficients, indicating the lack
of rich-club ordering and the lack of page-club ordering, i.e. high
in-degree (PageRank) pages purposely avoid sharing links with other
high in-degree (PageRank) pages. For UC patents citation network,
we identify page-club and rich-club ordering providing a conclusion
that {}``elite'' patents are cited by other {}``elite'' patents.
Finally, for e-mail communication network we show lack of both rich-club
and page-club ordering. We construct an example of synthetic network
showing page-club ordering and the lack of rich-club ordering. \end{abstract}
\maketitle

\section{Introduction}

The study of complex systems pervades through almost all the sciences,
from cell biology to ecology, from computer science to meteorology,
to name just a few. A paradigm of a complex system is a network, described
usually as a graph, where complexity may come from different sources:
topological structure, network evolution, connection and node diversity,
and/or dynamical evolution. Perhaps the most widely known graph property
is the \textit{node degree distribution} $P(k)$, which specifies
the probability of nodes having degree $k$ in a graph. The unexpected
findings that degree distributions of some real-world network topologies
closely follow power laws stimulated further interest in network research
\cite{books}.

However, node degree distribution does not describe the interconnectivity
of nodes with given degrees, that is, it does not provide any information
on the total number $m(k_{1},k_{2})$ of links between nodes of degree
$k_{1}$ and $k_{2}$. \textit{Joint degree distribution} is defined
as $P(k_{1},k_{2})=m(k_{1},k_{2})\mu(k_{1},k_{2})/(2m)$, where $\mu(k_{1},k_{2})$
is 2 if $k_{1}=k_{2}$ and 1 otherwise, and $m$ is the number of
links in the graph. Clearly joint degree distribution contains more
information about connectivity in a graph than degree distribution:
it provides information about 1-hop neighborhoods around a node. Given
$P(k_{1},k_{2})$, we can calculate $P(k)=({\bar{k}}/k)\sum_{k'}P(k,k')$,
but not vice versa, where ${\bar{k}}=\sum_{k}kP(k)$.

Although looking into the high order distributions is a complex task,
reminding us there is a price to pay, a well chosen set of metrics
can give us a simple partial view of high-order distributions. Several
such graph metrics that exploit joint degree distribution are: 
\begin{itemize}
\item \textit{Assortativity coefficient}: \[
r\sim\sum_{k_{1},k_{2}}^{k_{max}}k_{1}k_{2}\left[P(k_{1},k_{2})-\frac{k_{1}k_{2}P(k_{1})P(k_{2})}{{\bar{k}}^{2}}\right]\]

\item \textit{Average neighbor connectivity}: \[
k_{nn}(k)=\sum_{k'}^{k_{max}}k'P(k'|k)\]

\item \textit{Local clustering}: \[
C(k)=2m_{nn}(k)/[k(k-1)],\]
 where $m_{nn}(k)$ is the number of links between the neighbors of
$k$-degree nodes 
\item \textit{Rich-club coefficient}: \[
\phi(k)=\frac{2E_{>k}}{N_{>k}(N_{>k}-1)},\]
 where $E_{>k}$ is the number of edges among the $N_{>k}$ nodes
having degree higher than a given value $k$. 
\end{itemize}
In this paper we define two new metrics for directed graphs: rich-club
coefficient and page-club coefficient. These metrics can give us a
deeper level of understanding the complex networks as they represent
different projections of the joint degree distribution. When network
is undirected both metrics reduce to rich-club coefficient for undirected
graphs. The rich-club phenomenon refers to the tendency of nodes with
high centrality to form tightly interconnected communities. Since
this phenomenon is one of the crucial properties accounting for the
formation of dominant communities in real networks and since many
real networks are directed, in this paper we suggest a generalization
of this phenomenon to directed networks. The two metrics, rich-club
and page-club coefficients, may not be correlated: we construct an
example of synthetic (directed) network showing the page-club ordering
and the lack of rich-club ordering.

This is the outline of the paper. Section \ref{sec-pre} overviews
rich-club coefficient for undirected networks. In section \ref{sec-new}
two new metrics, rich-club and page-club coefficients, for directed
graphs are introduced. In section \ref{sec-real} these new metrics
are computed for 4 networks: (A) journal papers citation network,
(B) web graph (released from Google), (C) UC patents citation network,
and (D) e-mail communication network. Our conclusions are presented
in section \ref{sec-con}.

\section{Rich-club coefficient for undirected networks}

\label{sec-pre}

Graphs considered in this section are undirected and unweighted simple
graphs. The rich-club coefficient, introduced by Zhou and Mondragon
in the context of the Internet \cite{rich-club}, refers to the tendency
of high degree nodes, the hubs of the network, to be very well connected
to each other. Denoting by $E_{>k}$ the number of edges among the
$N_{>k}$ nodes having degree higher than a given value $k$, the
rich-club coefficient is expressed as: \begin{equation}
\phi(k)=\frac{2E_{>k}}{N_{>k}(N_{>k}-1)}\label{eq:rich-club}\end{equation}
 After some basic analytical analysis of the rich-club coefficient
\cite{detecting-rich-club}, we can see that it can be expressed
as a function of the joint degree distribution \begin{equation}
\phi(k)=\frac{N\left\langle k\right\rangle \sum_{k'=k+1}^{k_{max}}\sum_{k''=k+1}^{k_{max}}P(k',k'')}{N_{>k}(N_{>k}-1)}\label{eq:rich-club-high}\end{equation}
giving us a partial view of the high-order degree distribution which
is far more economical to compute.

In \cite{rich-club}, the rich-club coefficient $\phi$ is defined
in terms of nodes with rank less than $r_{max}$ where nodes are sorted
by decreasing degree values and the node rank $r$ denotes the position
of a node on this ordered list normalized by the total number of nodes.
Several networks are compared and a threshold value of 1\%, i.e. the
value of $\phi(1\%)$ was used to differentiate the networks and provide
evidence of the rich-club phenomenon. However, a monotonic increase
of $\phi(k)$ does not necessarily imply the presence of the rich-club
phenomenon. Indeed, even in the case of the ER graph -- a completely
random network -- has an increasing rich-club coefficient. This implies
that the increase of $\phi(k)$ is a natural consequence of the fact
that vertices with large degree have a larger probability of sharing
edges than low degree vertices. This feature is therefore imposed
by construction and does not represent a signature of any particular
organizing principle or structure, as is clear in the ER case \cite{detecting-rich-club}.
The simple inspection of the $\phi(k)$ trend is therefore potentially
misleading in the discrimination of the rich-club phenomenon, it can
only be used as a simple statistical property to differentiate several
networks in their complex structure.

Therefore, in order to detect rich-club phenomenon several null models
were proposed that normalize the basic rich-club coefficient. A null
model was presented in \cite{detecting-rich-club} where the rich-club
is normalized by the expression $\rho(k)=\phi(k)/\phi_{ran}(k)$ where
\begin{equation}
\phi_{ran}(k)=\frac{1}{N\left\langle k\right\rangle }\left[\frac{\sum_{k'=k+1}^{k_{max}}k'P(k')}{\sum_{k'=k+1}^{k_{max}}P(k')}\right]^{2}{\sim\atop k,k_{max}{\rightarrow\infty}}\frac{k^{2}}{\left\langle k\right\rangle N}\end{equation}
 is the rich-club coefficient of the maximally random network (uncorrelated
network) with the same degree distribution $P(k)$ as the network
under study. Operatively, the maximally random network can be thought
of as the stationary ensemble of networks visited by a process that,
at any time step, randomly selects a couple of links of the original
network and exchange two of their ending points (automatically preserving
the degree distribution)\cite{detecting-rich-club}. An actual rich-club
ordering is denoted by a ratio $\rho(k)>1$. Note that in sufficiently
large networks and large $k$, $\phi_{ran}(k)$ becomes clearly dependent
of $k$.

\begin{figure}
\center \resizebox{0.75\textwidth}{!}{\includegraphics{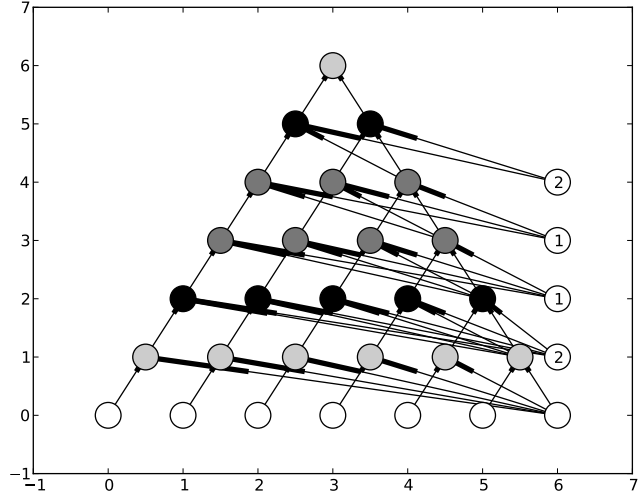}}
\caption{$d=6.$ Nodes are colored gradually according to their in-degree with
white to black color denoting lowest to highest in-degree respectively.
The additional nodes, i.e. nodes of the set $S_{i}$ are aggregated
in one node with label $\left|S_{i}\right|$ for simplicity.}

\label{Flo:synth-graph} 
\end{figure}

\begin{figure}
\center (a) \resizebox{0.75\textwidth}{!}{\includegraphics{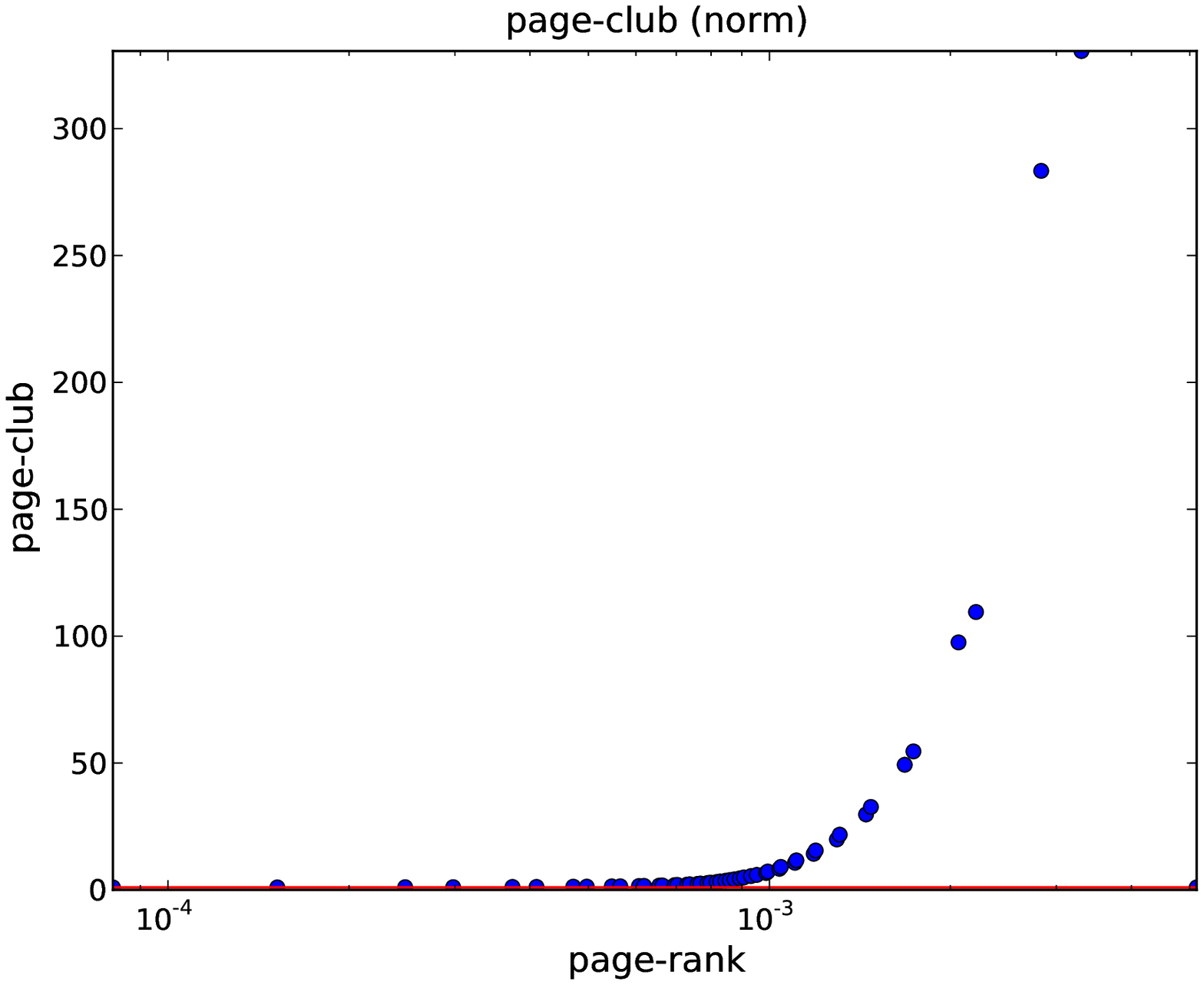}}
\\
 (b) \resizebox{0.75\textwidth}{!}{\includegraphics{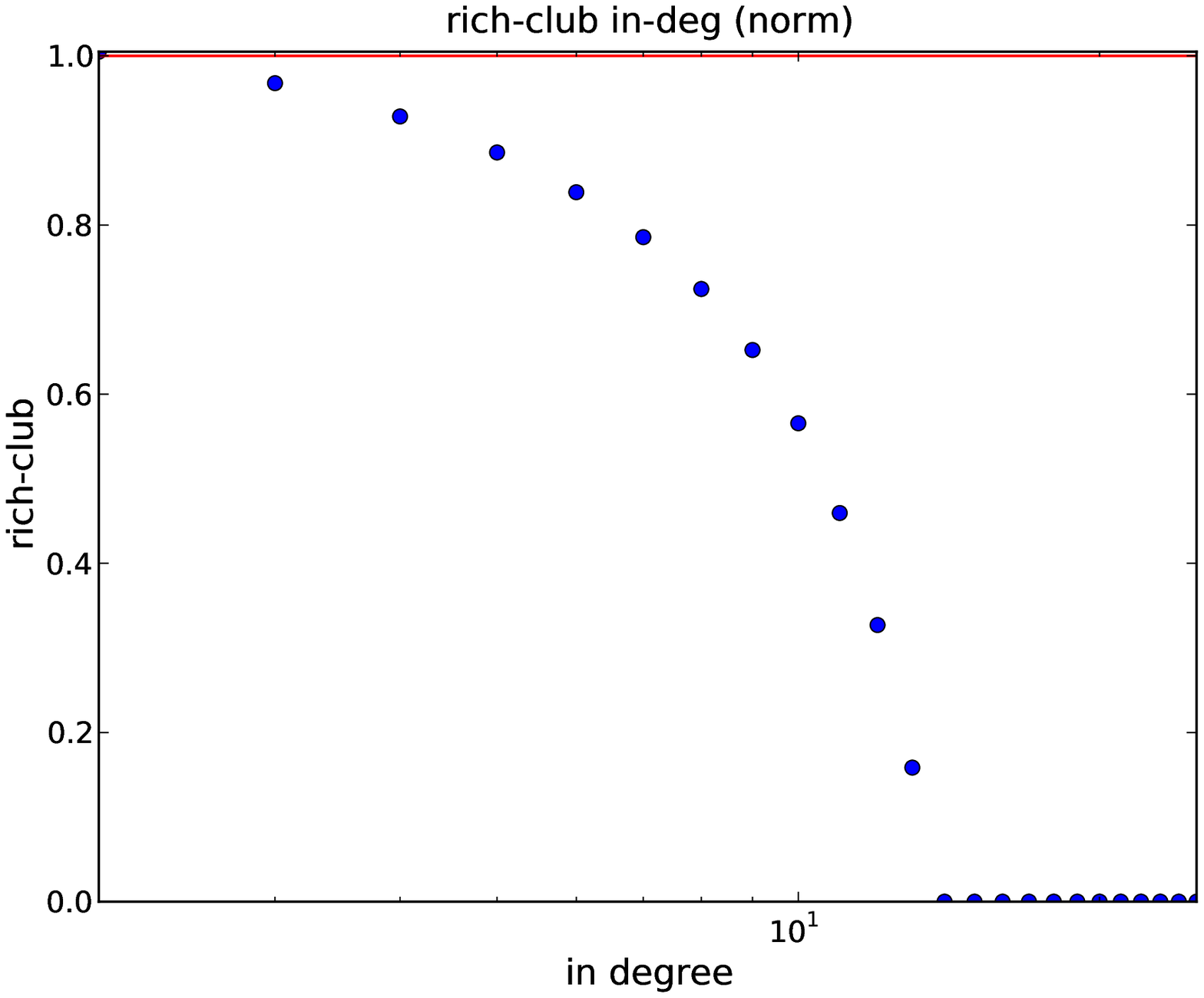}}
\caption[synthetic]{Synthetic network: (a) normalized page-club coefficient $\phi^{PR}(l)/\phi_{ran}^{PR}(l)$
versus page rank (b) normalized rich-club coefficient $\phi^{in}(k)/\phi_{ran}^{in}(k)$
versus in-degree. The network shows the lack of rich-club ordering,
but strong page-club ordering. }

\label{Flo:synth} 
\end{figure}

\section{Novel metrics for directed graphs}

\label{sec-new}

In this section we consider directed graphs. A directed graph (or
digraph) is a pair $G=(V,E)$ of a set $V$, whose elements are called
vertices or nodes, and a set $E$ of ordered pairs of vertices, called
arcs or directed edges.

\subsection{In-degree rich-club coefficient}

Having directed networks in mind, rich-club coefficient can be defined
in two ways, in terms of in-degree and out-degree. The one that we
are interested in is the in-degree rich-club coefficient defined,
in a very similar way to (\ref{eq:rich-club}), as %
\begin{equation}
\phi^{in}(k)=\frac{E_{>k}^{in}}{N_{>k}^{in}(N_{>k}^{in}-1)},\label{eq:rich-club-in-degree}\end{equation}
 where $E_{>k}^{in}$ is the number of directed edges among the $N_{>k}^{in}$
nodes having in-degree higher than a given value $k$. Note the number
2 missing in the numerator since in directed full-mesh graph the number
of edges is twice than that in the undirected graph.

We can express the numerator in (\ref{eq:rich-club-in-degree}) as
\begin{equation}
E_{>k}^{in}=\sum_{k'=k+1}^{k_{in}^{max}}\sum_{k''=k+1}^{k_{in}^{max}}E_{k'\rightarrow k''}^{in},\label{eq:e>k-rich-club}\end{equation}
 where $k_{in}^{max}$ is the maximum node in-degree of the network
and $E_{k'\rightarrow k''}^{in}$ denotes the number of edges pointing
from a node of in-degree $k'$ to a node of in-degree $k''$. Only
in the case of random uncorrelated networks, $E_{k'\rightarrow k''}^{in}$
takes the simple form \begin{equation}
E_{k'\rightarrow k''}^{in}=\frac{NP_{in}(k'')k''\left\langle k{}_{out}^{k_{in}=k'}\right\rangle P_{in}(k')}{\left\langle k_{in}\right\rangle },\label{eq:ek->k'-rich-club}\end{equation}
where $\left\langle k{}_{out}^{k_{in}=k'}\right\rangle $ denotes
the out-degree averaged over all nodes of in-degree $k'$ and $P_{in}(k)$
denotes the probability of a node having in-degree $k$. At first
sight $\left\langle k{}_{out}^{k_{in}=k'}\right\rangle $ may seem
constant in the case of large networks representing web graphs, but
having in mind the power-law distribution of in-degree, the number
of nodes belonging to the same in-degree class for high in-degree
becomes considerably small and is insufficient for converging $\left\langle k{}_{out}^{k_{in}=k'}\right\rangle $
to the general $\left\langle k{}_{out}\right\rangle $. By inserting
(\ref{eq:ek->k'-rich-club}) and (\ref{eq:e>k-rich-club}) into (\ref{eq:rich-club-in-degree})
we obtain the null model $\phi_{ran}^{in}(k)$ for uncorrelated directed
networks as \begin{eqnarray*}
\phi_{ran}^{in}(k) & = & \frac{N\sum_{k+1}^{k_{in}^{max}}k''P_{in}(k'')\sum_{k+1}^{k_{in}^{max}}\left\langle k_{out}^{k_{in}=k'}\right\rangle P_{in}(k')}{\left\langle k_{in}\right\rangle N_{>k}^{in}(N_{>k}^{in}-1)}\end{eqnarray*}

\begin{figure}
\center (a) \resizebox{0.75\textwidth}{!}{\includegraphics{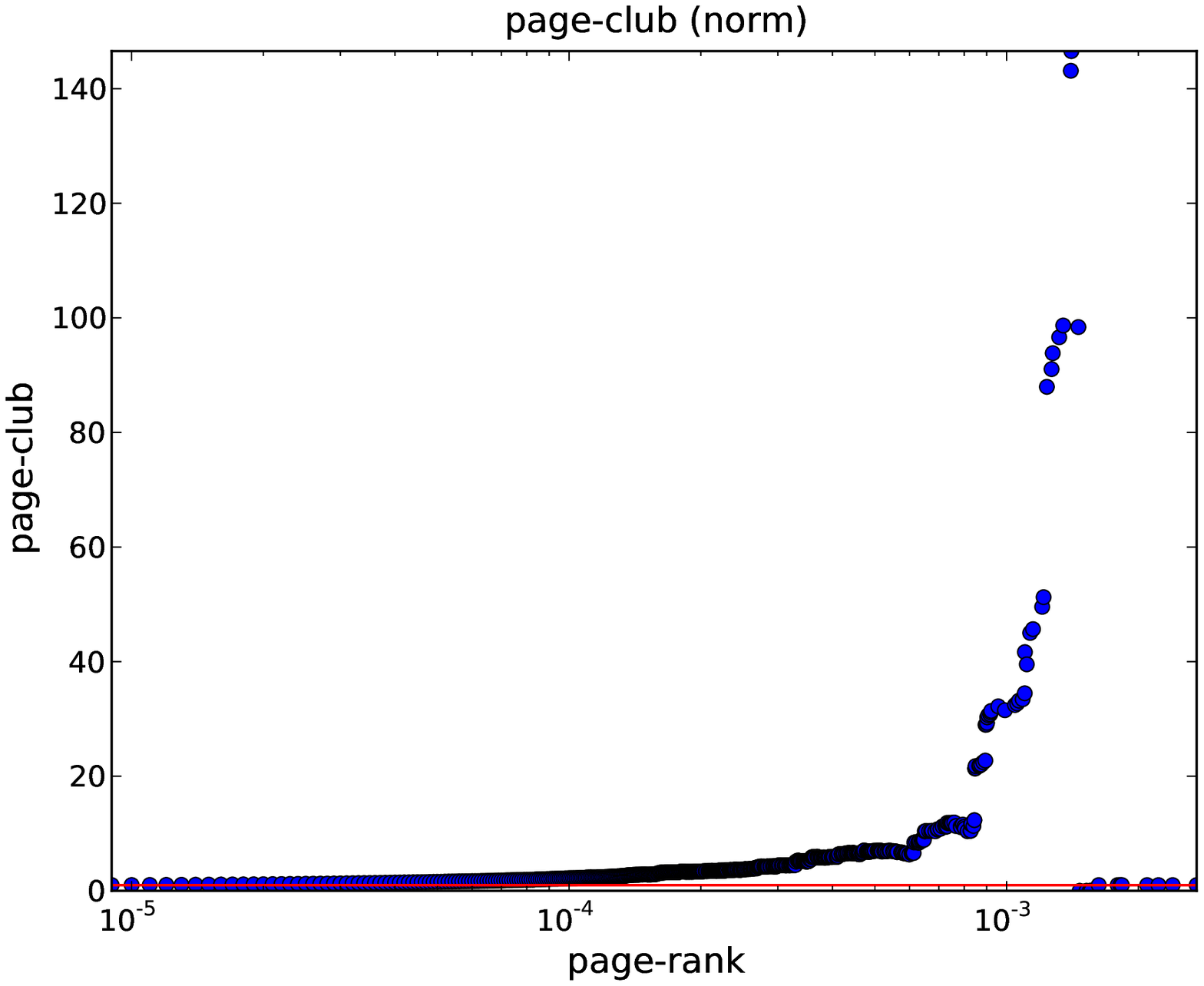}}
\\
 (b) \resizebox{0.75\textwidth}{!}{\includegraphics{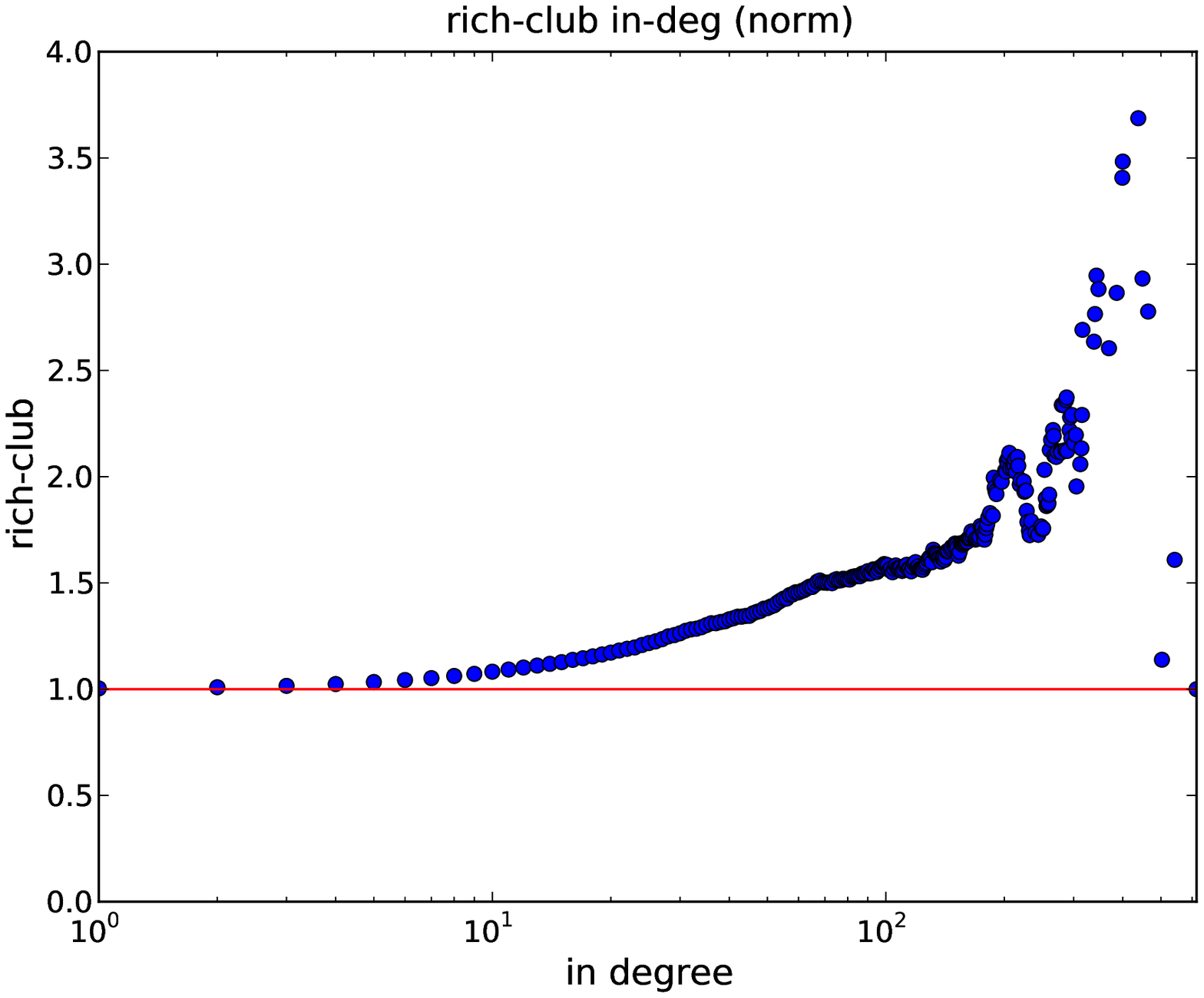}}
\caption[cit-jour]{Journal papers citation network: (a) normalized page-club coefficient
$\phi^{PR}(l)/\phi_{ran}^{PR}(l)$ versus page rank (b) normalized
rich-club coefficient $\phi^{in}(k)/\phi_{ran}^{in}(k)$ versus in-degree.
The network shows both page-club and rich-club ordering. Note that
the page-club ordering is much more stronger than the rich-club ordering. }

\label{Flo:cit-journal-2} 
\end{figure}

\subsection{Page-club coefficient}

Analogously to rich-club coefficient we define a new graph metric
called page-club coefficient which refers to the tendency of high
PageRank nodes, i.e. most popular pages in the web, to be highly interconnected.
Consider a modified random walker whose behavior is defined by the
following two rules: (a) with probability $1-q$, the walker follows
any outgoing link of $i$, chosen with equal probability, and (b)
with probability $q$ it moves to a generic node of the network (including
$i$), chosen with equal probability. Therefore, \[
pr(i)=\frac{q}{N}+(1-q)\sum_{j\in V;j\to i}\frac{pr(j)}{k_{out}(j)}\]
 We assume that each node has at least one outgoing link, and therefore
the last equation is well defined. Thus, for each $i$, stationary
probability $pr(i)$, called also PageRank, is well defined and $pr(i)>0$.
The probability $q$ is referred as damping factor; the damping factor
adopted in real applications is generally small ($q\approx0.15$).

Denoting by $E_{PR>l}$ the number of directed edges among the $N_{PR>l}$
nodes having PageRank value higher than a given value $l$, the page-club
coefficient is expressed as: \begin{equation}
\phi^{PR}(l)=\frac{E_{PR>l}}{N_{PR>l}(N_{PR>l}-1)}\label{eq:page-club}\end{equation}
Please note that in the uncorrelated networks this metric converges
to classical in-degree rich-club, since in uncorrelated networks the
average PageRank of nodes of the same in-degree class becomes linearly
dependent of the in-degree \cite{pr-in-degree}, i.e. \[
\overline{pr}(k_{in})=\frac{q}{N}+\frac{1-q}{N}\frac{k_{in}}{\left\langle k_{in}\right\rangle }\]
 where $\overline{pr}(k_{in})$ is the average PageRank of nodes with
in-degree $k_{in}$. Also, it is important to mention that relative
fluctuations of PageRank within the same class decrease as the in-degree
increases. Analogous to (\ref{eq:rich-club-in-degree}), an appropriate
null model for page-club can be defined \begin{eqnarray*}
\phi_{ran}^{PR}(l) & = & \frac{N}{\left\langle k_{in}\right\rangle N_{PR>l}(N_{PR>l}-1)}\times\\
 &  & {\sum_{l+1}^{l_{max}}\left\langle k_{in}^{pr=l''}\right\rangle P_{pr}(l'')\sum_{l+1}^{l_{max}}\left\langle k_{out}^{pr=l'}\right\rangle P_{pr}(l')}\end{eqnarray*}
where $P_{pr}(l)$ denotes the probability of a node having PageRank
$l$, and $\left\langle k_{in}^{pr=l}\right\rangle $ and $\left\langle k_{out}^{pr=l}\right\rangle $
are average node in-degree and average node out-degree, averaged over
all nodes of PageRank class $l$.

\subsection{Synthetic network}

Generally, all the results of the networks used in this paper show
a high correlation between page-club and rich-club coefficients, see
section \ref{sec-real}, but one should not derive a general conclusion
from this observation. In this section we generate a synthetic network
showing increase of page-club coefficient and decrease of rich-club
coefficient. Consider the directed tree graph $G=(V,E)$ where $V$
represents the node set, and $E$, the edge set. Let this tree be
with depth $d$, where the depth of a node $i$ is the length of the
path from the root to the node, and the depth of the tree is the maximal
length of all such paths. Further, let $G$ be a labeled graph, where
each node can have one of several labels (depending on its depth in
the tree), i.e. $V=V_{0}\cup V_{2}\ldots\cup V_{d}$. We denote by
$\left|V_{i}\right|$ the number of depth $i$ nodes. Assume that
$\left|V_{i}\right|=i+1,i=0,\ldots,d$, i.e. the number of nodes increments
as we move to a larger depth in the tree. We compose the edge set
of the ordered pairs of vertices $\{(V_{ik},V_{i-1,k})\mid k\in(1,\left|V_{i-1}\right|),i\in(0,d)\}\cup\{V_{i,\left|V_{i}\right|}\times V_{i-1}\mid i\in(0,d)\}$
where $V_{i,k}$ denotes the $k$-th node in $V_{i}$. In other words,
depth $i$ nodes propagate their PageRank score to depth $i-1$ nodes,
and furthermore, all nodes in the same class (depth) have equal PageRank
values. By this construction, all the nodes, except the leaves, have
in-degree 2. To change this property, we add additional set of nodes
$S_{i}$ connecting to the set $V_{i}$ with the edge set $E_{i}=S_{i}\times V_{i}$,
i.e. we increment the in-degree of the nodes in the set $V_{i}$ by
$\left|S_{i}\right|$. Note that the in-degree of these additional
nodes is 0. So, we tweak the in-degree of the nodes in the class $V_{i}$,
$k_{V_{i}}^{in}$, by the following rule: beside the leaf nodes, we
start with a $k_{V_{0}}^{in}=2$ and increment the in-degree by one
in every even depth, whereas for odd depths, we start with a high
in-degree in lower depths and decrement the in-degree as we go in
higher depths. A more formal definition would be \begin{equation}
k_{V_{i}}^{in}=\left\{ \begin{array}{cc}
0, & i=d\\
\frac{i}{2}+2, & i=2n,n\in\mathbb{N}\\
\left\lfloor \frac{d-i}{2}\right\rfloor +2, & i=2n+1,n\in\mathbb{N}\end{array}\right.\end{equation}

Such graph with depth $d=6$ is shown in Fig. \ref{Flo:synth-graph}.
What we want to achieve is the lack of rich-club ordering where nodes
with high in-degree connect to nodes with low in-degree and vise versa.
Further, a direct consequence of the tree structure is that nodes
with high page-rank will propagate their score to the successor nodes
and therefore positive page-club ordering should arise. The results
are shown in Fig. \ref{Flo:synth} for a generated graph of depth
50 with 1926 nodes. We observe the lack of rich-club ordering, but
strong page-club ordering, as expected. Also, we stress that the top
612 in-degree nodes are not sharing any links. Thus, for such networks,
the results of the analysis of the inter-connectivity of nodes, would
clearly depend on the definition of the {}``rich'' nodes (in-degree
or PageRank).

\begin{figure}
\center (a) \resizebox{0.75\textwidth}{!}{\includegraphics{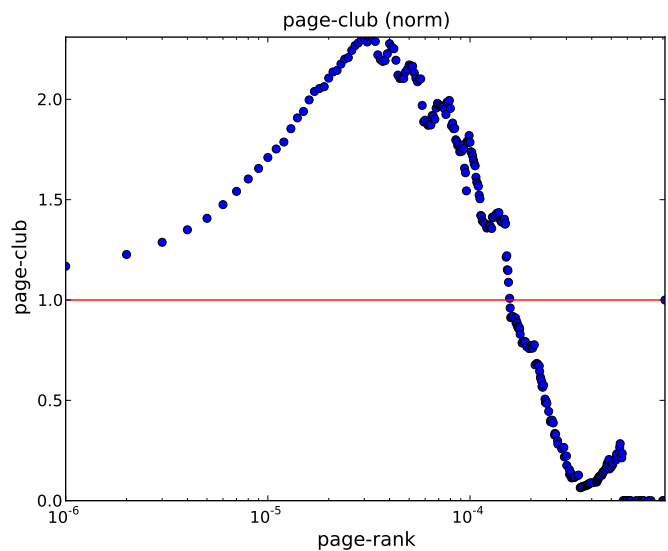}}
\\
 (b) \resizebox{0.75\textwidth}{!}{\includegraphics{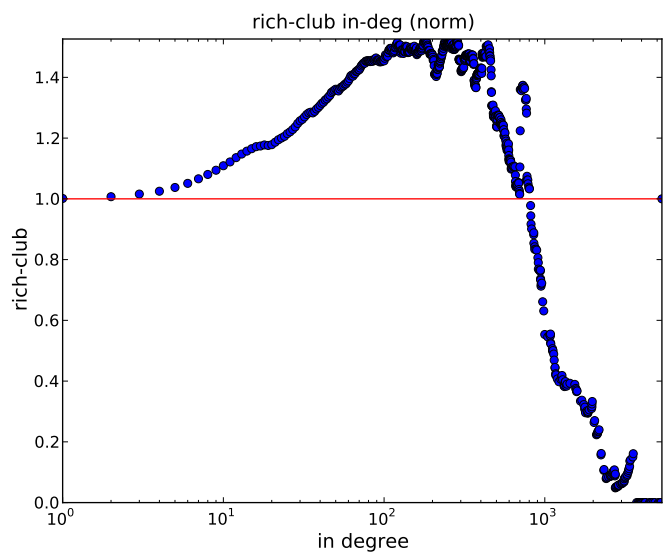}}
\caption[web-g]{Web graph (released in 2002 by Google): (a) normalized page-club
coefficient $\phi^{PR}(l)/\phi_{ran}^{PR}(l)$ versus page rank, and
(b) normalized rich-club coefficient $\phi^{in}(k)/\phi_{ran}^{in}(k)$
versus in-degree. The page-club and rich-club ordering are observed
only to some point, then the network shows the lack of page-club ordering
and the lack of rich-club ordering.}

\label{Flo:web-google-2} 
\end{figure}

\begin{figure}
\center (a) \resizebox{0.75\textwidth}{!}{\includegraphics{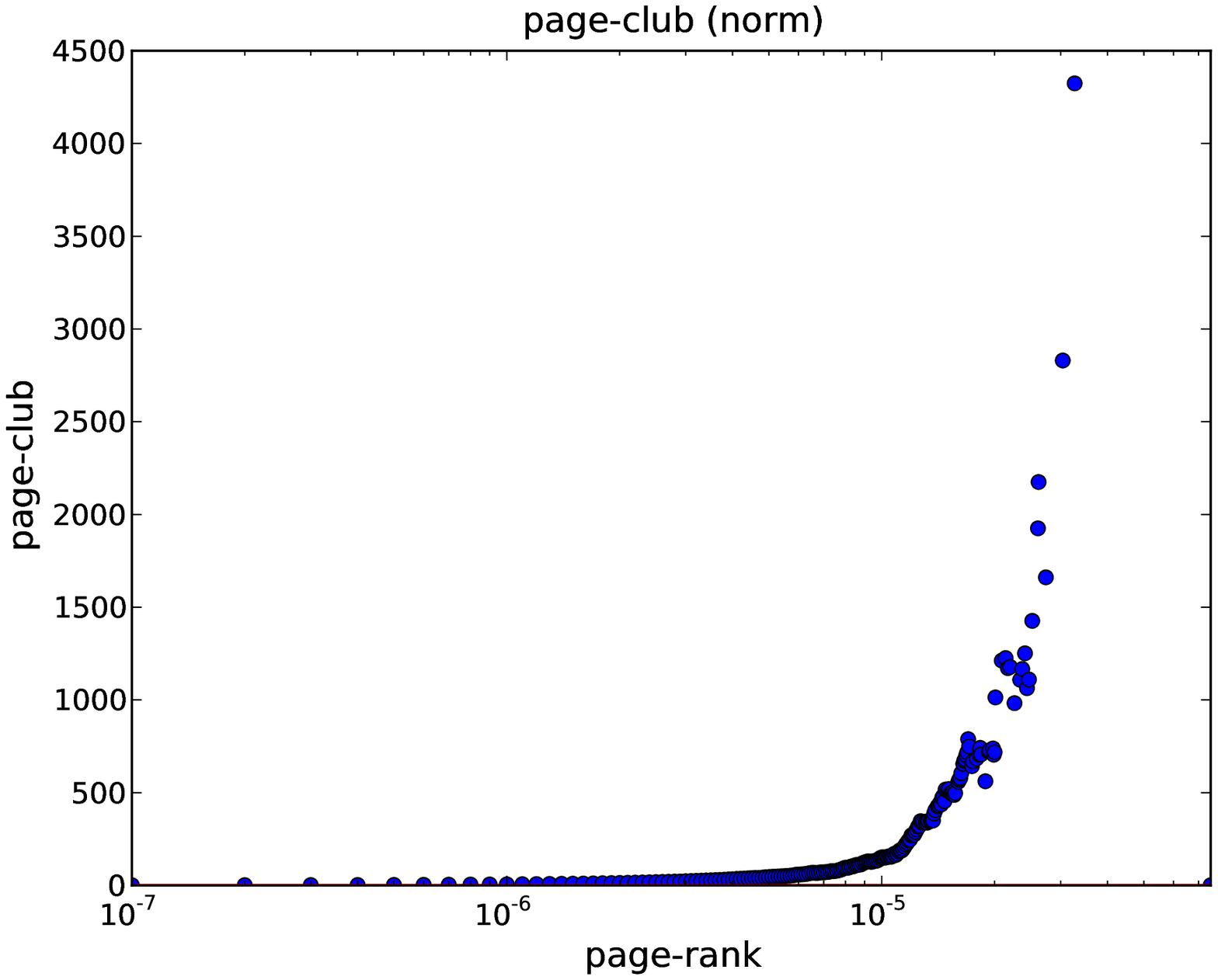}}
\\
 (b) \resizebox{0.75\textwidth}{!}{\includegraphics{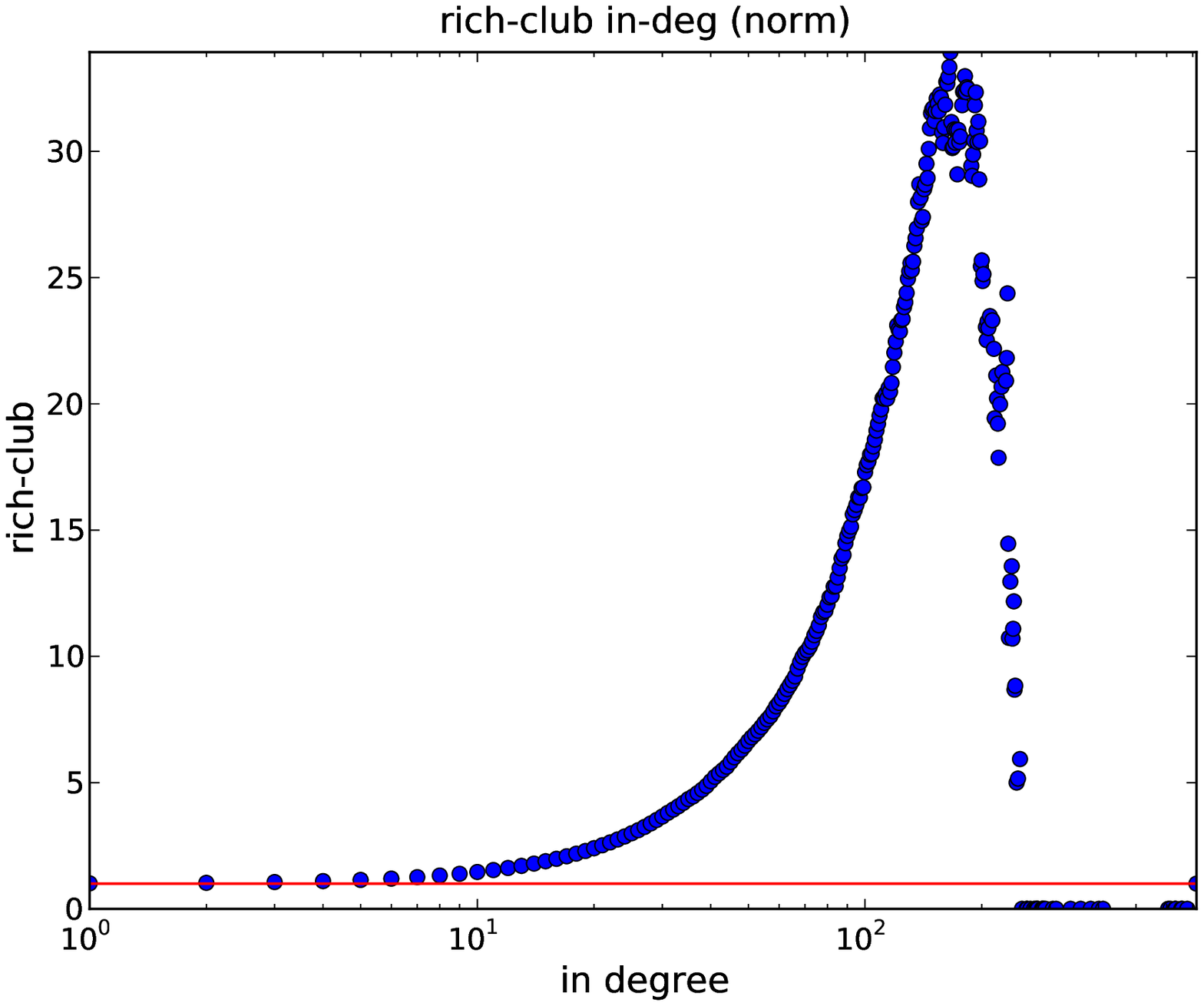}}
\caption[cit-pat]{US Patents citation network: (a) normalized page-club coefficient
$\phi^{PR}(l)/\phi_{ran}^{PR}(l)$ versus page rank (b) normalized
rich-club coefficient $\phi^{in}(k)/\phi_{ran}^{in}(k)$ versus in-degree.
The network shows both page-club and rich-club ordering. Note that
the page-club ordering is much more stronger than the rich-club ordering. }

\label{Flo:cit-patents-2} 
\end{figure}


\section{Real networks: results and discussions}

\label{sec-real}

The network data used in this paper consists of four networks \cite{jure-snap}: 
\begin{itemize}
\item Arxiv High Energy Physics paper citation network (cit-HepPh): Directed,
Temporal, Labeled network with 34,546 nodes and 421,578 edges; 
\item Web graph from Google (web-Google): Directed network with 875,713
nodes and 5,105,039 edges; 
\item Citation network among US Patents (cit-Patents): Directed, Temporal,
Labeled network with 3,774,768 nodes and 16,518,948 edges; and 
\item Email network from a EU research institution (email-EuAll): Directed
network with 265,214 nodes and 420,045 edges. 
\end{itemize}
For each network we compute normalized rich-club and page-club coefficients
$\rho_{in}(k)=\phi^{in}(k)/\phi_{ran}^{in}(k)$, $\rho_{PR}(l)=\phi^{PR}(l)/\phi_{ran}^{PR}(l)$.
We stress that $\rho_{in}(k)$ and $\rho_{PR}(l)$ may, in some cases,
be undefined. In the following, we discuss only the quantity $\rho_{in}(k)$
since the discussion for $\rho_{PR}(l)$ is exactly same. $\rho_{in}(k)$
is undefined when its denominator is equal to zero, $\phi_{ran}^{in}(k)=0$.
We rewrite $\phi_{ran}^{in}(k)$ as \begin{equation}
\phi_{ran}^{in}(k)=\frac{inlinks_{>k}outlinks_{>k}}{\left|E\right|N_{>k}^{in}(N_{>k}^{in}-1)}\end{equation}
 where $inlinks_{>k}$ ($outlinks_{>k}$) denotes all the \emph{in-links}
(\emph{out-links}) arriving to (departing from) nodes that have in-degree
greater than $k$ and $\left|E\right|$ denotes the number of directed
edges in the network.

We consider several cases: 
\begin{itemize}
\item When $N_{>k}^{in}=1$ or $N_{>k}^{in}=0$, i.e. when we have a single
node or no nodes in the {}``club''. 
\item When $inlinks_{>k}=0$. Note that this case should not happen in practice
since in the two special cases of $\phi_{ran}^{in}(0)$ and $\phi_{ran}^{in}(k_{in}^{max})$
we generally have a positive number of \emph{in-links}. 
\item When $outlinks{}_{>k}=0$. This case happens in tree graphs, such
as citation networks, where the top in-degree nodes (roots) have no
\emph{out-links}. 
\end{itemize}
We handle all these cases by assigning $\rho_{in}(k)$ a value 1.
It is important to stress that $\rho_{in}(k)$ can have a value zero,
i.e. its denominator can be well defined, thus having a positive number
of \emph{out-links} departing the {}``club'', but no links are shared
within the {}``club''.

We also stress that for the PageRank computation we used the damping
factor $q=0.15$ for all the networks. We also used $q=0.5$ for the
journal citation network as proposed by \cite{pagerank-citation}
but no significant changes are observed, so these results are omitted.

\begin{figure}
\center (a) \resizebox{0.75\textwidth}{!}{\includegraphics{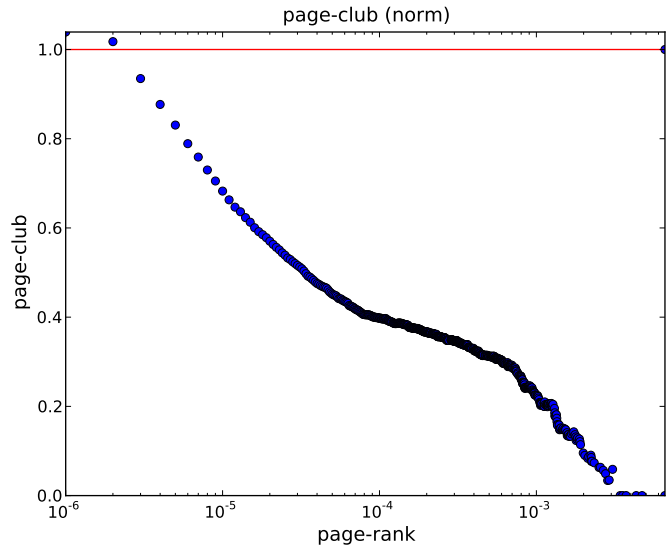}}
\\
 (b) \resizebox{0.75\textwidth}{!}{\includegraphics{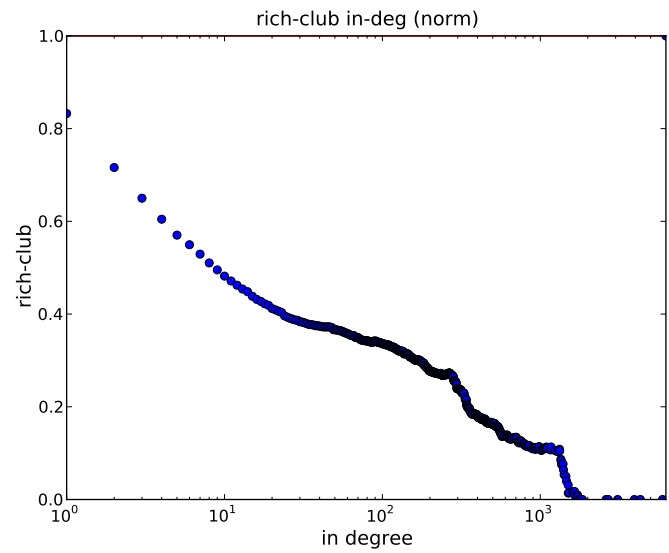}}
\caption[comm-email]{Email communication network: (a) normalized page-club coefficient
$\phi^{PR}(l)/\phi_{ran}^{PR}(l)$ versus page rank (b) normalized
rich-club coefficient $\phi^{in}(k)/\phi_{ran}^{in}(k)$ versus in-degree.
The lack of page-club ordering and the lack of rich-club ordering
are observed. }

\label{Flo:comm-email-2} 
\end{figure}

\subsection{Papers citation network}

The Arxiv High Energy Physics paper citation network is formed from
the e-print arXiv dataset and covers all the citations within a dataset
of 34,546 papers with 421,578 edges. If a paper $i$ cites paper $j$,
the graph contains a directed edge from $i$ to $j$. If a paper cites,
or is cited by, a paper outside the dataset, the graph does not contain
any information about this. The data covers papers in the period from
January 1993 to April 2003 (124 months). It begins within a few months
of the inception of the arXiv, and thus represents essentially the
complete history of its HEP-PH section. The graph has an exponential
degree distribution and a tree structure.

In Fig. \ref{Flo:cit-journal-2} we show the normalized rich-club
and page-club coefficients and identify both rich-club and page-club
ordering providing a conclusion that {}``elite'' papers are cited
by other {}``elite'' papers. If we say that {}``elite'' papers
are written by {}``elite'' scientists and those scientists decide
to reference {}``elite'' papers, i.e. papers written by other {}``elite''
scientists than this result coincides with previous findings \cite{detecting-rich-club}
i.e. it indicates existence of an {}``oligarchy'' of highly influential
and mutually communicating scientists. Note the difference between
page-club and rich-club. The page-club ordering is much more stronger
than the rich-club which can be explained by having in mind the tree
structure in citation network, i.e. older papers which are higher
in the hierarchy have higher PageRank retrieved from all the successors
independently of the number of their direct successors, i.e., their
in-degree. We also point that the top 8 PageRank nodes (roots) have
no out-links, thus having a page-club value of one, whereas the top
in-degree nodes have a positive rich-club ordering.

\subsection{Google Web graph}

In the web graph released in 2002 by Google as a part of Google Programming
Contest, nodes represent web pages and directed edges represent hyperlinks
between them. The graph consists of nearly one million (1,000,000)
nodes and over five million (5,000,000) edges following a power-law
degree distribution.

In Fig. \ref{Flo:web-google-2} we show the normalized rich-club and
page-club coefficients where we identify partial rich-club (page-club)
ordering up to some point (the middle layer) and then a narrow declining
of the coefficients, which show the lack of rich-club ordering and
the lack of the page-club ordering. In other words, in this networks
high-degree (PageRank) pages purposely avoid sharing links with other
high-degree (PageRank) pages. Some sort of competitiveness among strong
pages could be a possible explanation of this phenomenon. Also note
the uncorrelated property of the network, therefore explaining the
high similarity between page-club and rich-club coefficient \cite{pr-in-degree}.
We stress that the top 24 (20) in-degree (PageRank) nodes are not
sharing any links between them beside their positive number of out-links,
therefore the zero values of the rich-club and page-club coefficients.

\subsection{US Patents citation network}

The U.S. patent dataset, maintained by the National Bureau of Economic
Research, spans 37 years (January 1, 1963 to December 30, 1999), and
includes all the utility patents granted during that period, totaling
about four million (4,000,000) patents. The citation graph includes
all citations made by patents granted between 1975 and 1999, totaling
16,522,438 citations. For the patents dataset there are 1,803,511
nodes for which we have no information about their citations (we only
have the in-links).

In Fig. \ref{Flo:cit-patents-2}b we show the normalized rich-club
coefficient: the rich-club ordering is observed up to some point,
and after that point, the ordering quickly decreases to zero, where
the top 27 in-degree nodes are not sharing any links between them.
Fig. \ref{Flo:cit-patents-2}a shows the normalized page-club coefficient:
one could identify page-club ordering providing the conclusion that
top PageRank patents are cited by top PageRank patents. Note the much
stronger page-club than rich-club ordering, generally, because of
the tree structure.

\subsection{Email communication network}

The network of email communication of a large European research institution
contains all incoming and outgoing email of the research institution
for the period of October 2003 to May 2005 (18 months). Given a set
of email messages, each node corresponds to an email address. A directed
edge between nodes $i$ and $j$ was created if $i$ sent at least
one message to $j$. The network consists of 265214 nodes and 420045
edges.

Fig. \ref{Flo:comm-email-2} shows the normalized rich-club and page-club
coefficients. The lack of the page-club ordering and the lack of the
rich-club ordering for this network could be explained by observing
that scientists are working in research groups where each group has
one to few {}``elite'' scientists managing the group, where communication
between the {}``elite'' scientists from different groups is reduced
to a minimum. The top 9 (6) in-degree (PageRank) nodes are not sharing
any links between them beside their positive number of out-links,
therefore the zero values of the rich-club and page-club coefficients.

\section{Conclusion}

\label{sec-con}

In this paper two new metrics for directed graphs are introduced,
namely the normalized rich-club coefficient and the normalized page-club
coefficient. For different directed graphs these two coefficients
are computed. The results have indicated a high correlation between
page-club and rich-club coefficients except for the synthetic network,
for which the coefficients have opposite behavior. In general, beside
the high correlation observed in several real networks, these metrics
are not same. Detecting rich-club phenomenon often used to indicate
the dominance of an {}``oligarchy'' of {}``rich'' and mutually
communicating entities. However, this analysis clearly depends of
the definition of {}``rich'' nodes. The page-club coefficient annotates
nodes with high PageRank as the {}``popular'' nodes, thus, in networks
where PageRank emerges as a natural metric for distinguishing between
popular and unpopular nodes, one should use page-club to indicate
the emergence of an {}``oligarchy'' formed by {}``elite'' nodes.

\section*{Acknowledgments}
We wish to gratefully acknowledge the support of EU project MANMADE (Grant No. 043363).

\bibliographystyle{elsarticle-num}

\end{document}